\documentclass[a4paper,10pt]{extarticle}

 \usepackage{epsfig}

\usepackage{amssymb}
\usepackage{bm}

\usepackage[utf8]{inputenc}
\usepackage[russian]{babel}
\usepackage{amsmath}
\usepackage{amsfonts} 
\usepackage{listings}
\usepackage{graphicx}
\usepackage{textcomp}
\usepackage{wrapfig}    
\usepackage{psfrag}
\usepackage{multicol}
\usepackage{caption}

\begin{document}
\begin{center}

\textbf{\large Статистические методы поиска локальных источников нейтрино высоких энергий 
в проекте <<Байкальский глубоководный нейтринный телескоп>>}


\bigskip
\bigskip
\bigskip
\bigskip

Т.А.Овсянникова и О.В.Суворова

\bigskip
для коллаборации Байкал
\bigskip
\bigskip

\begin{abstract}
Целью данной работы является исследование и применение методов поиска локальных астрофизических
источников нейтрино на глубоководных нейтринных телескопах НТ200 и НТ1000 в озере Байкал.
В отсутствии превышения измеренных событий над фоном для НТ200  за 1038 дней живого времени наблюдений
вычислены верхние пределы на число событий и на нейтринный поток от выбранных локальных 
источников гамма-излучения. Делается оценка чувствительности к тем же источникам для НТ1000 
за год наблюдений. Для этого исследуются два алгоритма -- биновый и небиновый способ нахождения 
скоплений в заданном направлении прихода частиц на установку. Их сравнение делается по величине 
статистической значимости. Среди подходов статистического анализа выбран метод
максимального правдоподобия.


\end{abstract}
\bigskip
\bigskip

\vspace{\fill}

Москва ИЯИ РАН -- \number\year

\end{center}


\newpage
\tableofcontents
\newpage

\section{Введение}\label{sec:levelIntro}

Обнаружение потоков нейтрино высоких и сверхвысоких энергий от астрофизических источников и понимание механизмов 
их генерации является актуальной фундаментальной проблемой, которая решается на современных нейтринных телескопах. 
Недавние результаты полярного телескопа IceCube~\cite{ICubePeV2013} по измерению энерговыделения и направления 
приходящих частиц впервые указывают на наблюдение событий от взаимодействия нейтрино внеатмосферного происхождения. 
При этом показано, что для энергий ниже 60 тераэлектронвольт (ТэВ) энергетическое 
распределение событий соответствует ожидаемому фону от атмосферных мюонов и нейтрино, а для (28) событий с энергией 
выше 60 ТэВ нулевая гипотеза (только фон) отрицается на уровне 4.1 сигма~\cite{ICubePeV2013}. Вместе 
с тем угловое распределение
этих событий соответствует изотропному. С учетом того, что новый Байкальский проект НТ1000~\cite{GVD-NT1000}
будет иметь эффективный объем 400 Мтонн, такой же как в IceCube для событий с фиксированной вершиной взаимодействия, 
и примерно в 4 раза лучшее угловое разрешение для каскадных событий~\cite{GVD-NT1000} чем в IceCube, то в поиске
астрофизических нейтрино актуальным остается метод распознавания их скоплений на небесной сфере.  

Источниками нейтрино с энергиями, превышающими сотни гигаэлектронвольт (ГэВ), рассматриваются астрофизические объекты, 
наблюдаемые в гамма лучах, такие как пульсары, остатки сверхновых, микроквазары, 
активные ядра галактик, где предполагаются механизмы ускорения частиц, например, вследствие ударных волн 
(Ферми-ускорение) при их распространении в облаках намагниченной плазмы. В результате реакций столкновения 
ускоренных протонов с веществом или их рассеяния на фотонах радиационных полей в окрестности источника
рождаются нейтральные и заряженные $\pi$-мезоны, от распада которых генерируются потоки фотонов и нейтрино
высоких энергий. В соответствии с механизмом Ферми-ускорения 
энергетическое распределение протонов от астрофизических объектов имеет зависимость $\propto E^{-2}$.
Максимальные энергии протонов, до которых  они могут быть ускорены в таких системах, порядка $10^{19}$ 
электронвольт (эВ) и максимальные энергии в генерированных 
потоках нейтрино примерно такой же величины. 

Основными фоном в поиске локальных источников нейтрино на подводных телескопах
являются потоки мюонов и нейтрино атмосферного происхождения т.е. от распада пионов и каонов, 
генерированных во взаимодействиях космических лучей (а это в основном протоны) с ядрами воздуха. 
Критерии отбора событий из нижней полусферы должны уменьшать эти потоки на 6--7 порядков в 
соответствии с методом (предложенным М.А. Марковым в 1960 году) глубоководного детектирования 
взаимодействия нейтрино с веществом в реакциях глубоко неупругого рассеяния: 
по черенковскому излучению инициированных релятивистских мюонов и каскадов (адронных или электромагнитных) 
при их распространении в прозрачной среде.
Более того из-за крутого наклона энергетического спектра $\propto E^{-3}$ атмосферных нейтрино
основная доля этих событий в наблюдении относится к интервалу энергий от 100 МэВ до 100 ТэВ, и увеличение порога 
регистрации существенно снижает их число. Предполагается, что для пороговых энергий от 30 ТэВ конфигурация НТ1000 
позволит эффективно отбирать астрофизические нейтрино по каскадным событиям 
с частотой не менее 10 событий в год~\cite{GVD-NT1000}. 

Экспериментальная база данных Байкальского глубоководного нейтринного телескопа 
НТ200~\cite{NT200-1, NT200-2, NT200-3},
с эффективным объемом регистрации нейтрино порядка ста килотонн, 
накоплена за пять лет со времени запуска в апреле 1998 года на глубине погружения 1100 м. 
В месте расположения телескопа НТ200 с 2011 года по-этапно реализуется новый Байкальский 
проект НТ1000 гигатонного размера. Его первый из двенадцати кластеров оптических модулей (ОМ)
планируется запустить в апреле 2015 года. Структура каждого кластера соответствует увеличенной 
по высоте (345 м) и радиусу (60 м) модели НТ200 из 8 струн со 192 ОМ (15 м между ОМ на струне)
и эффективному объемому $\sim$10 мегатонн. 
Оценка чувствительности установок Байкальского проекта к потоку астрофизических нейтрино остается актуальной.
Первые результаты поиска локальных источников нейтрино на телескопе НТ200 представлены в работе~\cite{NT200-AstroBelolap2013}
методом бинового анализа. Другие подходы в анализе данных по поиску источников нейтрино применены
на телескопах международных коллабораций IceCube~\cite{ICube-2011}, ANTARES~\cite{ANTARES-2012}, Super Kamiokande~\cite{SuperKam-2009}.

Целью данной работы является реализация методик поиска
локальных источников по нейтринным событиям телескопа НТ200 за 1038 дней
живого времени наблюдений, а также оценка чувствительности НТ1000 за год к потоку нейтрино 
от выбранных локальных источников гамма-излучения.  Препринт организован следующим образом.
В разделе~\ref{sec:levelStat} дается определение статистических величин и критериев оценок, которые
используются в анализе. В разделе~\ref{sec:levelNT200} описывается небиновый подход анализа данных
и сравнивается с биновым. В разделе~\ref{sec:levelUpL} приведен расчет пределов НТ200 и чувствительности
НТ1000 к потоку астрофизических нейтрино от 42 предполагамых локальных источников. В разделе~\ref{sec:levelSummary}
приведены основные выводы сделанного анализа.

\section{Статистические методы анализа данных}\label{sec:levelStat}
Можно выделить две основные задачи в статистическом анализе данных: оценку
параметров измеренного распределения и проверку гипотез о законе распределения измеренных величин. 
Оценка параметров c использованием непараметрических методов
в случаях подобных измерению природных потоков нейтрино,
когда нет возможности получить истинные повторности наблюдений, оказывается предпочтительнее
параметрического. В этом подходе нет априорного предположения о законе распределения изучаемой 
случайной величины. При этом широко используется метод генерации повторных выборок, например, алгоритмом бутстреп.
Основная идея этого метода состоит в том, чтобы из конечной совокупности N измерений
с помощью генератора случайных чисел, равномерно распределенных на интервале [1, N], 
выбирать произвольный элемент исходной выборки на каждом шаге из N последовательных итераций. 
Таким способом формируется любое число псевдовыборок и по ним определяются 
необходимые параметры эмпирического распределения: оценки математического ожидания, дисперсии, 
доверительного интервала или делаются проверки его приближенной аппроксимации выбранным 
теоретическим распределением. С увеличением N с полученными значениями статистик можно обращаться 
как с независимыми случайными величинами. Как будет показано далее, мы использовали этот метод для
определения фоновых событий в измеренном наборе данных НТ200. 

Известны два подхода в поиске сигнала относительно фона: биновый и небиновый методы анализа. 
В биновом способе выбор размера бина и числа бинов ("сетка") делается c условиями 
либо равной площади бинов, либо равного фона на бин, либо одинаковой эффективности сигнала на бин. 
Пространственно-временные координаты нейтринных событий для определения бинов
преобразуются в сферические экваториальные координаты: склонение и прямое восхождение.
Для каждого бина проверяется гипотеза о законе распределения измеренных событий и
находится верхний предел на число сигнальных событий на заданном доверительном уровне.
В небиновом методе фиксируются центры сетки, относительно которых в заданных
радиусах области фоновых событий и области для поиска сигнала проверяются 
нулевая гипотеза (что измерен только фон) и альтернативная построением
функций максимального правдоподобия. Небиновым методом возможна более точная
локализация сигнальной области.

Мы используем тестирование отношения правдоподобия в проверке гипотез,
а в определении верхних пределов метод Байеса, предполагая, что 
число измеренных и число фоновых событий подчиняются статистике Пуассона. 
Результат сравнивается с альтернативным (частотным) подходом, где
одновременно учитываются систематические погрешности эксперимента и теории.
Ниже приведены аналитические выражения для этих вычислений.

\subsection{Метод функции максимального правдоподобия}
Метод заключается в том, чтобы получить максимальную вероятность воспроизвести экспериментальный результат, 
предполагая некоторую теоретическую зависимость.
Если задана функция плотности вероятности $f(x_i,\alpha_i)$, зависящая от N переменных $x_i$ и параметров $\alpha_i$, 
то вероятность найти величину $x_i$ в интервале $[x_i,x_i+dx_i]$ есть~\cite{bPDG}:
$$ dP=\prod\nolimits_{i=1}^N f(x_i,\alpha_i)dx_i.$$
Тогда функция максимального правдоподобия это есть произведение функций плотности вероятностей $L(\alpha_i)= \prod\nolimits_{i=1}^N f(x_i,\alpha_i)$. 
Чтобы определить максимальную вероятность dP, надо найти максимум функции $L(\alpha_i)$
дифференцированием по параметру $\alpha_i$:
$$\frac{\partial L}{\partial(\alpha_i)}=0~~~\Rightarrow \alpha^F_i.$$
На практике находят минимум $\mathfrak{L}= -Ln(L)$, что эквивалентно максимуму L.

Можно показать, что оценка одного стандартного отклонения вычисляется 
как $1~\sigma = -Ln(L) + 0.5$ и двух сигма как $2~\sigma = -Ln(L) + 2.$

\subsection{Байесовский подход}
В Байесовском подходе все знания о параметре $\alpha$ базируются на апостериорной вероятности, то есть вероятности 
иметь данное значение параметра при условии, что оно имеет заданное распределение. Вероятность определяется 
с помощью теоремы Байеса~\cite{bPDG}:
$$p(\vec \alpha|\vec x)= \frac{p(\vec x|\vec \alpha)p(\vec \alpha)}{p(\vec x)}=\frac{L(\vec x|\vec \alpha)\times\pi(\vec \alpha)}{\int L(\vec x|\vec \alpha)\times \pi(\vec \alpha)},$$
где $L(\vec x|\vec \alpha)$- функция максимального правдоподобия, а $\pi(\vec \alpha)$ -- априорная плотность вероятности
распределения параметра $\vec \alpha$.
Априорная вероятность показывает уровень доверия для параметра $\vec \alpha$ до того, как были произведены измерения.
Сложности в определении априорной вероятности привели к применению так называемой ``объективной байесовской статистики”, 
где априорная вероятность основана не на уровне доверия, а полученна из формальных правил. 
Например, когда априорная вероятность инвариантна относительно
преобразования параметров или соответствует максимальному количеству информации для данного набора измерений.
Априорные вероятности, вообще говоря, не отражают степень доверия, но они могут в некоторых
случаях рассматриваться как субъективные априорные вероятности. Апостериорные
вероятности тоже не обязательно отражают степень доверия. Объективные априорные вероятности вместе
с теоремой Байеса используются для оценок или доверительных интервалов, которые отвечают желательным свойствами.
Процедура вывода априорной объективной вероятности основана на правиле Джеффрис~\cite{bByesFreq}:
$$ \pi(\vec \alpha)\propto \sqrt(det (I(\vec \alpha))),$$
где
$$ I_{ij}(\vec \alpha)=-E\left[\frac{\partial^2 Ln L(\vec x|\vec \alpha) }{\partial \alpha_i~ \partial \alpha_j}\right]
= - \int \frac{\partial^2 Ln L(\vec x|\vec \alpha) }{\partial \alpha_i~ \partial \alpha_j} L(\vec x|\vec \alpha) d \vec x$$
есть информационная матрица Фишера~\cite{bByesFreq}. Такой подход приводит к инвариантности относительно преобразования параметров. 
Априорная вероятность Джеффрис 
содержит зависимость от функции правдоподобия, а следовательно, содержит информацию о модели. По этой причине
она не всегда нормируема на единицу. Однако при этом может нормироваться апостериорная вероятность, если
сама вероятность достаточно быстро падает. Насколько для данного набора независимых данных $\vec x$ 
априорная вероятность $\pi(\vec \alpha)$ согласуется с апостериорной вероятностью  $p(\vec x|\vec \alpha)$ 
показывает величина, называемая расстоянием Кульбака-Лейблера~\cite{bByesFreq}:
$$ D_n [\pi p] =\int p(\vec \alpha|\vec x)ln\frac{p(\vec \alpha|\vec x)}{\pi(\vec \alpha)} d\vec \alpha.$$\\
При этом зависимость от априорной вероятности выбрана так, что среднее значение этой величины 
максимально для предельного случая $n \rightarrow \infty$. Среднее значение апостериорной вероятности
$p(\vec \alpha|\vec x)$ вычисляется из распределения данных как~\cite{bPDG}:
$$ p(\vec x)= \int L(\vec x|\vec \alpha) \pi(\vec \alpha) d \vec \alpha.$$

\subsection{Верхний предел}
Когда цель эксперимента состоит в определении параметра $\alpha$, то требуется оценить некий интервал (C.L.), который
отражает статистическую значимость измерений. Если плотность вероятности не соответствует распределению Гаусса 
или если существуют физические ограничения на возможное значение параметра, то обычно устанавливается нижний и 
верхний предел на это значение, который должен покрывать истинное значение параметра заданной вероятностью и
заключать выводы о параметрах, относящихся к заявленной априорной вероятности.
При достаточно большой выборке данных, точечная оценка и стандартное отклонение удовлетворяют,
по существу, этим целям. 

Верхний предел ($\alpha_{up}$) в Байесовском подходе входит в уравнение как верхний предел интеграла~\cite{bPDG}:
\begin{equation}
1-C.L.= \int^{\alpha_{up}}_{\alpha_{lo}} p(\alpha|x) d\alpha
\label{eq:CLup}
\end{equation}
В предположении, что вероятность распределена по Пуассону, верхний предел ($s_{up}$)
находится численным решением уравнения~\cite{bPDG, bByesFreq}:
\begin{equation}
1-C.L.= e^{-s_{up}}\frac{\sum^n_{m=1}(s_{up}+b)^m/m!}{\sum^n_{m=1} b^m/m!}~~~\Rightarrow s_{up} 
\label{eq:Suplim}
\end{equation}\\

\subsection{Включение систематических ошибок}
Байесовская методика обеспечивает основу для включения систематических ошибок в конечный результат. Если модель 
зависит не только от параметров представляющих интерес ($\alpha$), но и от несущественных параметров ($\nu$), 
значения которых известны с некоторой ограниченной точностью, то в функции правдоподобия апостериорная или 
априорная плотности вероятности распределения величины $\vec x$ будут зависеть как от $\alpha$ так и от $\nu$. 
По теореме Байеса можно получить апостериорную плотность вероятности для $\alpha$  
путем интегрирования по несущественному параметру ~\cite{bPDG}:
$$ p(\alpha|\vec x)= \int p(\vec \alpha,\vec \nu|\vec x) d \vec \nu.$$
Например, в случае распределения вероятностей по Пуассону $P(n|s+b)$ оценки систематических ошибок 
рассматриваются как стандартные отклонения, экпериментальные и теоретические, относительно своих средних 
b (фона) и s (сигнала), и интегрируются в Гауссовом распределении. Такой метод был предложен в 90-х
в работе~\cite{bBobCous}, он имеет свои ограничения в применении.

Отметим, что статистическая обработка данных в альтернативном частотном методе -- альтернативном 
Баейсовскому подходу, последнее время используется в анализе таких экспериментов как IceCube 
и ANTARES. Соответственно, ими используется частотный подход для нахождения доверительных интервалов 
с включением систематических ошибок в функцию правдоподобия 
способом, представленным в работе~\cite{bConrad}. Практическое применение этого метода доступно 
в среде ROOT выбором установок в классе TRolke~\cite{bROOT}.

\subsection{Проверка гипотез}
Предположение о распределении параметров набора экспериментальных данных является некой гипотезой. 
Например, гипотеза H0, часто называемая нулевой, предполагает, что экспериментальные свойства событий 
происходят от фоновых процессов, а альтернативная гипотеза H1 предполагает наличие кроме фоновых событий 
еще и сигнальных. Статистический критерий значимости, по которому гипотеза H0 может 
быть принята или отвергнута, определяется нахождением области $X$ для набора экспериментальных
данных $\vec x$, где вероятность попасть в эту область меньше, чем заданный доверительный уровень  ($C.L.$). 
Если данные находятся в этой области, то H0 отвергается и принимается альтернативная гипотеза H1. 
В результате проверки H0 относительно гипотезы H1 можно прийти к правильному решению, 
либо совершить одну из двух ошибок: опровергнуть H0, когда она верна (ошибка первого рода), 
либо принять ее когда верна H1 (ошибка второго рода). Вероятность $(1-C.L.)$ -- ошибка первого рода.

Чтобы определить значимость H0 гипотезы по отношению к альтернативной H1, используется отношение правдоподобия.
Область $X$ выбирается так, что для всех данных $\vec x$ внутри $X$, удовлетворяются соотношения~\cite{bPDG}
\begin{equation}
\lambda=\frac{L(\vec x|H1)}{L(\vec x|H0)} \le \eta,\\
\label{eq:Lyambda}
 \end{equation}
 где H0, H1 -- простые гипотезы, не зависящие от неопределенных параметров, а
 $\eta$ выбирают так, чтоб получить заданный доверительный уровень: 
 $$P(\lambda(\vec x) \le \eta|\vec x)=C.L.$$

\subsection{Критерий значимости}
Часто требуется определить  уровень согласия  между данными и гипотезой H0 без предположения альтернативной гипотезы. 
Это делается по выборке данных $\vec x$, статистика которых отражает уровень
согласия  между данными и гипотезой. 
Значимость расхождения между данными и того, что ожидалось в предположении гипотезы H0, дает значение p-value.
Другими словами, p-value -- это допустимый уровень значимости, при котором принимается H0~\cite{bPDG}: 
$$ p=\int^{ \infty}_{x_{obs}} g(x|H0) dx,$$
где $x_{obs}$ -- число событий, наблюдаемое в эксперименте, и $g(x|H0)$ -- плотность вероятности данной статистики.

Все рассмотренные в этом разделе~\ref{sec:levelStat} определения и статистические методы, далее
используются в анализе данных Байкальского нейтринного эксперимента.

\section{Поиск локальных источников в проекте Байкальский глубоководный телескоп}\label{sec:levelNT200}

\subsection{Эксперимент: распределение нейтринных событий на сфере}

\begin{figure}[h!]
\center{\includegraphics[width=0.5\linewidth]{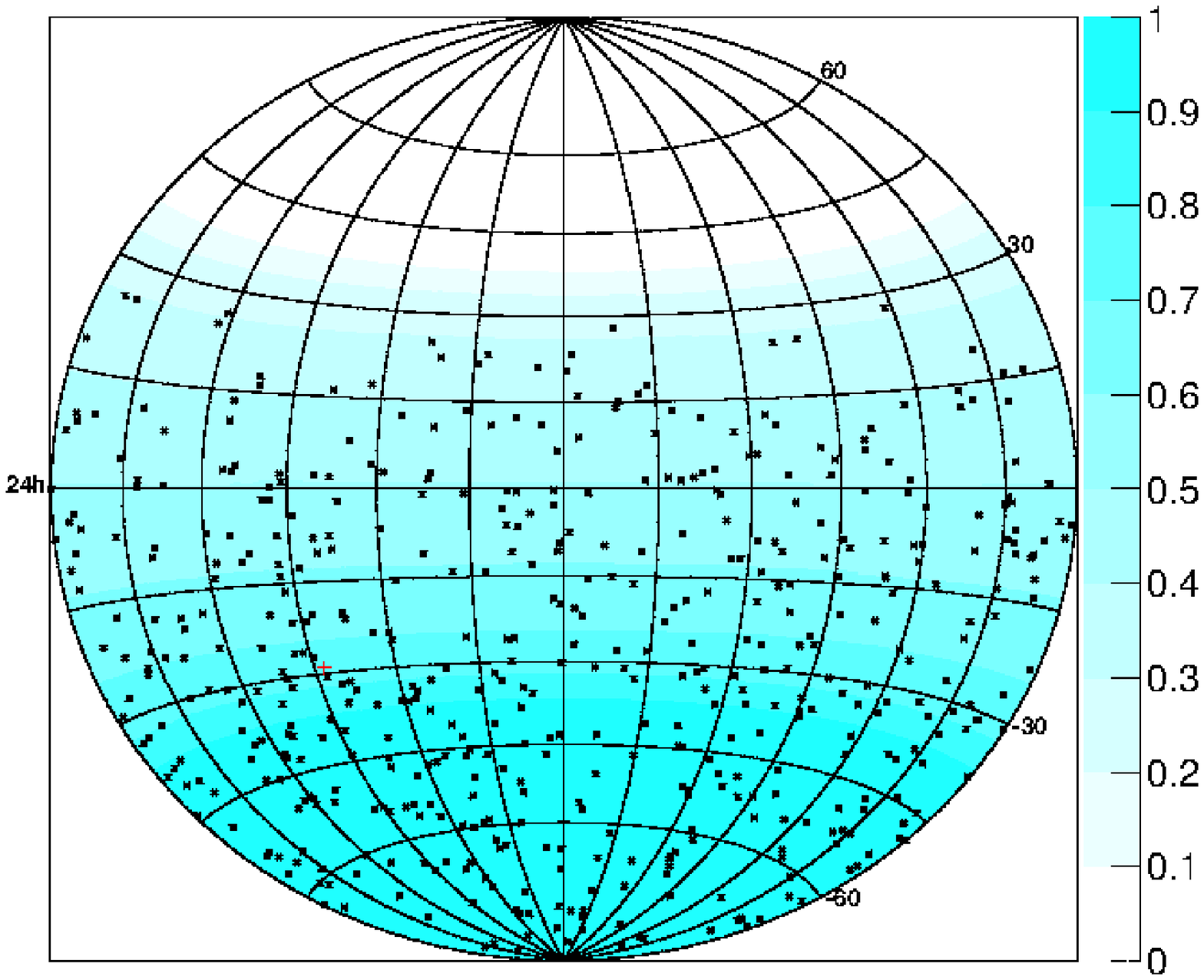} \\ }
\caption{\label{eqTvis} 
Видимая область неба в экваториальных координатах для Байкальского НТ200
и распределение измеренных нейтринных событий (черные ромбики). Положение центра Галактики
отмечено красным крестиком}
\end{figure}

Анализ экспериментальных данных телескопа НТ200 делается по выборке 396 нейтринных событий, отобранных 
по всем тем критериям выделения событий из-под горизонта, что были определены в работе ~\cite{bRusBelolap}, где
показано, что пороговая энергия мюонных нейтрино 15 ГэВ и угловое разрешение $4^{\circ}$. Координаты событий, измеренных в локальной горизонтальной 
системе, мы преобразуем в экваториальную в соответствии с формулами сферической астрономии ~\cite{AstroForm}:
\begin{equation}
\begin{split}
sin(\delta) = sin(\alpha) sin(\lambda) + cos(h) cos(\lambda) cos(A),    \\                               
cos(t) = ( sin(h) - sin(\lambda) sin(\delta) )/( cos(\lambda) cos(\delta) ), \\                              
t = LST- \alpha,
\end{split}
\label{eq:TranH2E}
\end{equation}
где LST -- местное звездное время, $\lambda=51.50^{\circ}$ -- географическая широта Байкальского телескопа. Здесь горизонтальные 
координаты это высота $h$ и азимут $A$, а экваториальные координаты  -- склонение $\delta$  и часовой угол t, 
либо склонение $\delta$  и восхождение $\alpha$, как следует из их определения. В пересчете координат использовалась 
также астрономическая библиотека подпрограмм slalib~\cite{SlaLib}. Напомним, что в экваториальной системе основной плоскостью является плоскость небесного экватора. Угол между плоскостью небесного экватора 
и направлением на источник называется склонением и отсчитывается от $0^{\circ}$ до $+90^{\circ}$ к северному 
полюсу мира и от $0^{\circ}$ до $-90^{\circ}$ к южному. Часовым углом t светила называется угол между плоскостями 
небесного меридиана и круга склонения светила. Часовые углы отсчитываются в сторону суточного вращения небесной 
сферы, то есть к западу от верхней точки небесного экватора, в пределах от $0^{o}$ до $360^{o}$ (в градусной мере) 
или от 0h до 24h (в часовой мере). Соответственно, восхождение  $\alpha$ 
это угол между направлением на точку весеннего равноденствия и плоскостью круга склонения светила, отсчитываются 
в сторону противоположную суточному вращению небесной сферы. Распределение измеренных нейтринных событий на сфере
в экваториальных $\delta$-$\alpha$ 
координатах показано на  Рис.~\ref{eqTvis} с градиентом (шкала цвета в долях) видимости областей неба в течение года 
для Байкальского нейтринного телескопа в соответствии с его широтой. В дальшейшем расчете предельных
потоков от предполагаемых локальных источников доля времени наблюдения каждого источника по зенитному углу
учитывалась как вес в интегрировании эффективной площади, зависящей от энергии нейтрино и его направления прихода. 

\subsection{Определение фона методом генерации псевдовыборок}
\begin{figure}[h!]
\begin{minipage}[h]{0.53\linewidth}
\center{\includegraphics[width=1\linewidth]{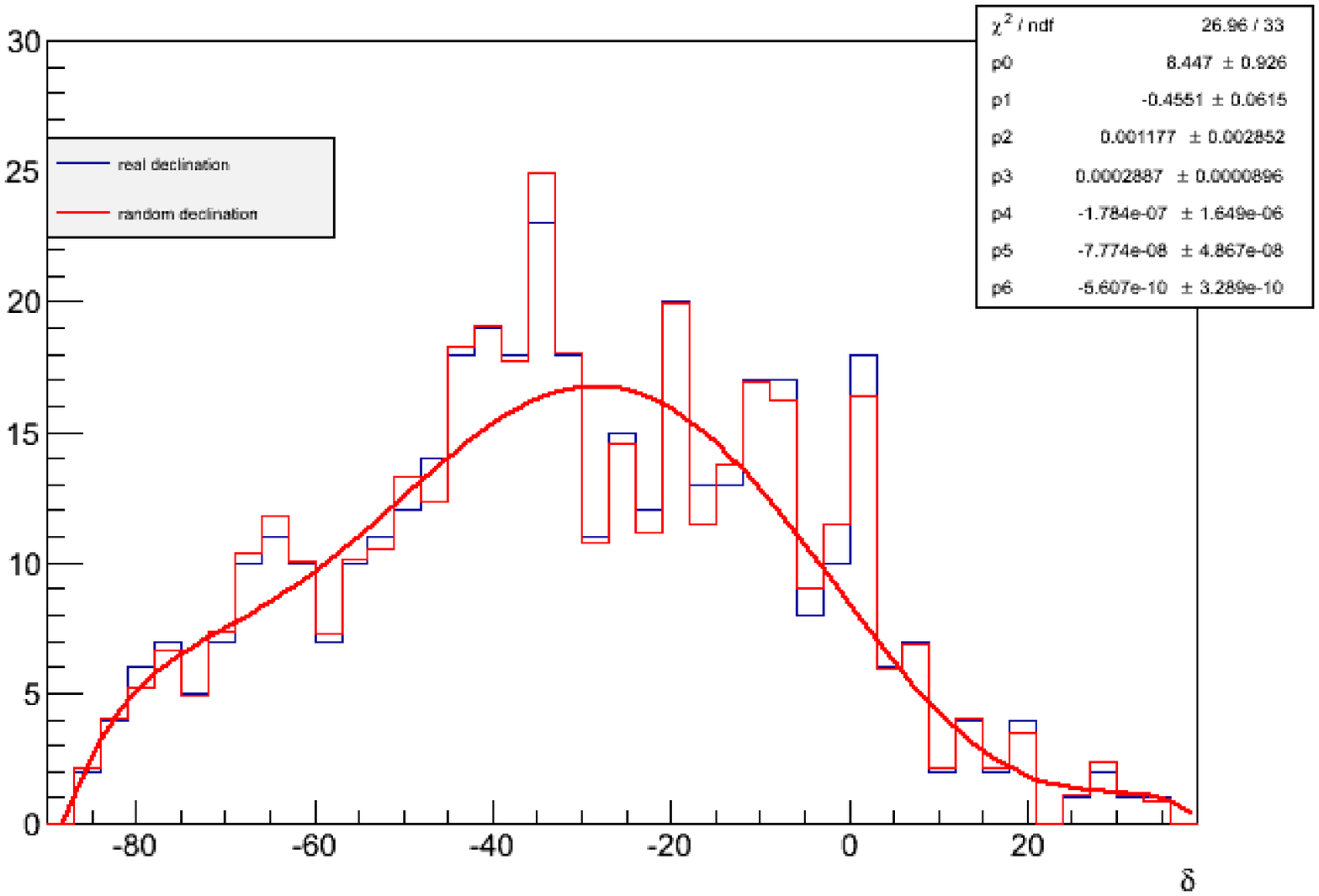} \\ }
\end{minipage}
\hfill%
\begin{minipage}[h]{0.53\linewidth}
\center{\includegraphics[width=1\linewidth]{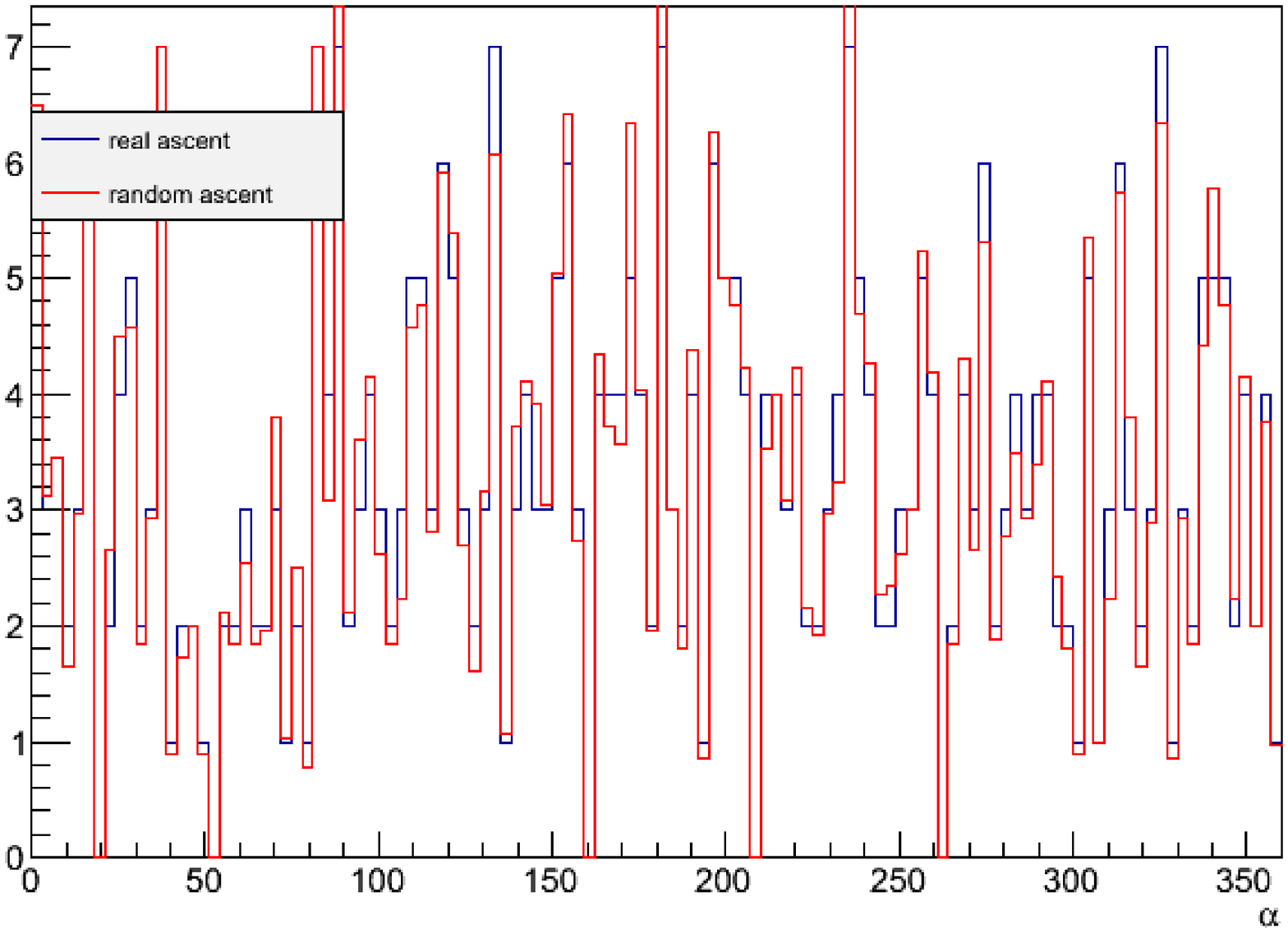} \\ }
\end{minipage}
\caption{\label{RandDec396} Распределения нейтринных событий НТ200 по склонению (слева) и 
прямому восхождению (справа):
синие гистограммы - реальные события, красные -- разыгранный фон. Красная линия это
функция, фитирующая распределение по склонению реальных событий.}
\end{figure}

\begin{figure}[h!]
\center{\includegraphics[width=0.7\linewidth]{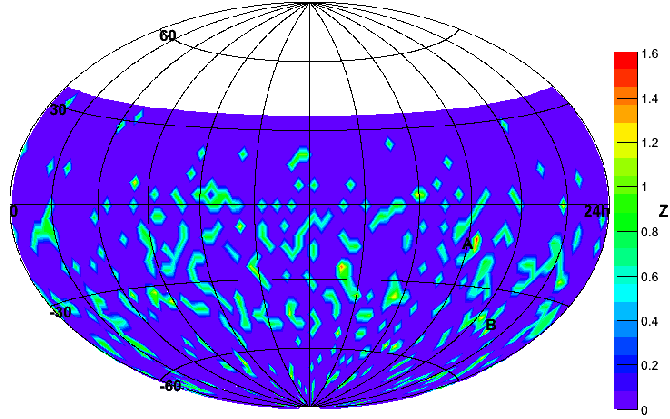} \\ }
\caption{\label{Z396} Значимость Z по данным НТ200 в биновом методе сканирования неба
в экваториальных координатах. Вероятное направление на скопление: А $(18.2h,-12^{\circ})$.} 
\end{figure}

Распределение фоновых событий в экваториальных $\delta$-$\alpha$ координатах
было получено методом бутсреп из двух распределений измеренных событий: по склонению и 
восхождению, и рассматриваемых как независимые наборы измерений. Для случайного перемешивания 
использовались генераторы в среде ROOT~\cite{bROOT}: TRandom и TRandom2, где последний имеет значительно больший период и 
поэтому был выбран для симуляций в конечном варианте. Как известно, по восхождению
фоновые события приходят равномерно, а по склонению - в зависимости от аксептанса
и эффективности телескопа. Для генерации псевдовыборок по восхождению сравнивалась
схема (А) равномерного розыгрыша в интервале значений 0-360 градусов и схема (В) равномерного
розыгрыша по номеру события для последующего выбора восхождения. Для генерации псевдовыборок по
склонению были рассмотрены схемы симуляций (А) по функции, фитирующей измеренное распределение,
и (В) по номеру события для последующего выбора склонения. Число иттераций было выбрано как 1000-кратное 
повторение полного числа нейтринных событий при заполнении "дерева" (класс TTree) координат фоновых
событий. Распределение реальных событий с фитирующей их функцией и распределение перемешанных событий
по схеме (В) приведены на Рис.~\ref{RandDec396} слева по склонению и справа по восхождению. 
\\
Для сравнения реальных и фоновых событий в экваториальных координатах в 2-х мерном распределении  
размер бина  был взят $ 4^\circ \times 4^\circ$. На Рис.~\ref{Z396} показано биновое распределение 
статистической значимости Z в проекции aitoff (в  ROOTе) в экваториальных координатах, где цвет 
соотносится со шкалой значений Z, вычисленных для каждого бина по формуле:
 \begin{equation}
 Z=\frac{N_{s}}{\sqrt{N_{s}+N_{b}}},
\label{eq:Zbin}
\end{equation}
где $N_{s}$ -- количество сигнальных событий, $N_{b}$ -- количество фоновых событий.
Как видно из Рис.~\ref{Z396}, указаний на статистически значимый избыток событий в каком-либо направлении нет. 
Тем не менее, наибольшая значимость здесь $ Z =1.6 $ соответствует координатам 
центра бина $(\alpha,\delta) =(18.2h,-12^{\circ})$, обозначенного на Рис.~\ref{Z396} символом А. 
Заметим, что в данном сравнении не было ограничения на минимальное число фоновых событий.

\subsection{Поиск скоплений событий с помощью небинового метода анализа}
В небиновом методе направления на небесной сфере выбирались сканированием по сетке в экваториальных 
координатах  $\delta$-$\alpha$ с шагом заданного радиуса:
от $3^\circ$ до $0.5^\circ$ для оценки зависимости конечного результата от размера шага.
Отбирались события, попавшие в конус радиусом $8^{\circ}$ вокруг исследуемого направления
для реального набора данных и в конус радиусом $20^{\circ}$ для фоновых событий.  
Для каждого направления рассчитывалось отношение правдоподобия $\lambda$ 
(см. уравнение (\ref{eq:Lambda})) и p-value, в предположении, что количество событий 
в конусе имеют распределение Пуассона. Функция правдоподобия составлялась следущим образом. 
Обозначим $\xi$ вероятность i-того события быть сигналом от точечного источника, тогда
($1-\xi$) это его вероятность быть фоном.
Плотность распределения вероятности для i-того события есть $$P_i=S(\theta_i)\times\xi + B(\theta_i)\times(1-\xi),$$ 
где $S(\theta_i)$ и $B(\theta_i)$ есть плотности вероятности сигнала и фона, соответственно, 
а $\theta_i$ -- пространственный угол между направлением на источник и событием.
В экваториальных координатах угол $\theta_i$ вычисляется как арккосинус выражения
$$ cos(\theta_i) =  sin(\delta_S) sin(\delta_i) + cos(\delta_S) cos(\delta_i) cos(\alpha_S - \alpha_i), $$
где $\delta_S$ и $\delta_i$ -- это склонения источника и события, а $\alpha_S$ и $\alpha_i$ -- прямые
восхождения, соответственно.\\ 
Вероятность попадания случайного события в телесный угол $\Omega$ это
$$dP=2\pi \frac{d\cos(\theta)}{\Omega}$$
и в нашем случае плотностью вероятности фоновых событий будет величина:
$$B(\theta_i)= 2 \pi \frac{\sin{\theta_i}}{\Omega}.$$
Распределение плотности вероятности для точечного источника задается распределением Гаусса: 
$$ S(\theta_i)= \frac{1}{2 \pi \sigma_{t}^2}e^{\frac{-|\theta_i|^2}{\sigma_{t}^2}},$$
где $\sigma_{t}$ -- угловое разрешение детектора. 
Угловое разрешение телескопа НТ200 бралось $4^{\circ}$, как в работе \cite{bRusBelolap}.
Функция максимального правдоподобия для данной задачи:
$$ L(\xi)=\prod\nolimits_{i=1}^N(S(\theta_i)\times\xi+B(\theta_i)\times(1-\xi)),$$
где N- число событий, наблюдаемое в данном конусе.
Оптимальное значение $\xi^F$ определялось, как описано в разделе~\ref{sec:levelStat}, 
минимизацией функции $-LnL(\xi)$. Поиск минимума функции $-LnL(\xi)$ делался численно, с помощью программного 
пакета ``MINUIT``, используя ''MIGRAD'', в среде ROOT, и далее оценивалось отношение правдоподобия:
\begin{equation}
\lambda=-2\log(\frac{L(\xi=0)}{L(\xi_F)}).
\label{eq:Lambda}
\end{equation} 

Отметим, что тестирование описанного алгоритма небинового метода было сделано в работе~\cite{TAO-Diplom2013} 
с данными каталога FERMI-LAT~\cite{bFERMI} для гамма источников с энергией выше ГэВ.

\begin{figure}[h!]
\begin{minipage}[h]{0.53\linewidth}
\center{\includegraphics[width=1\linewidth]{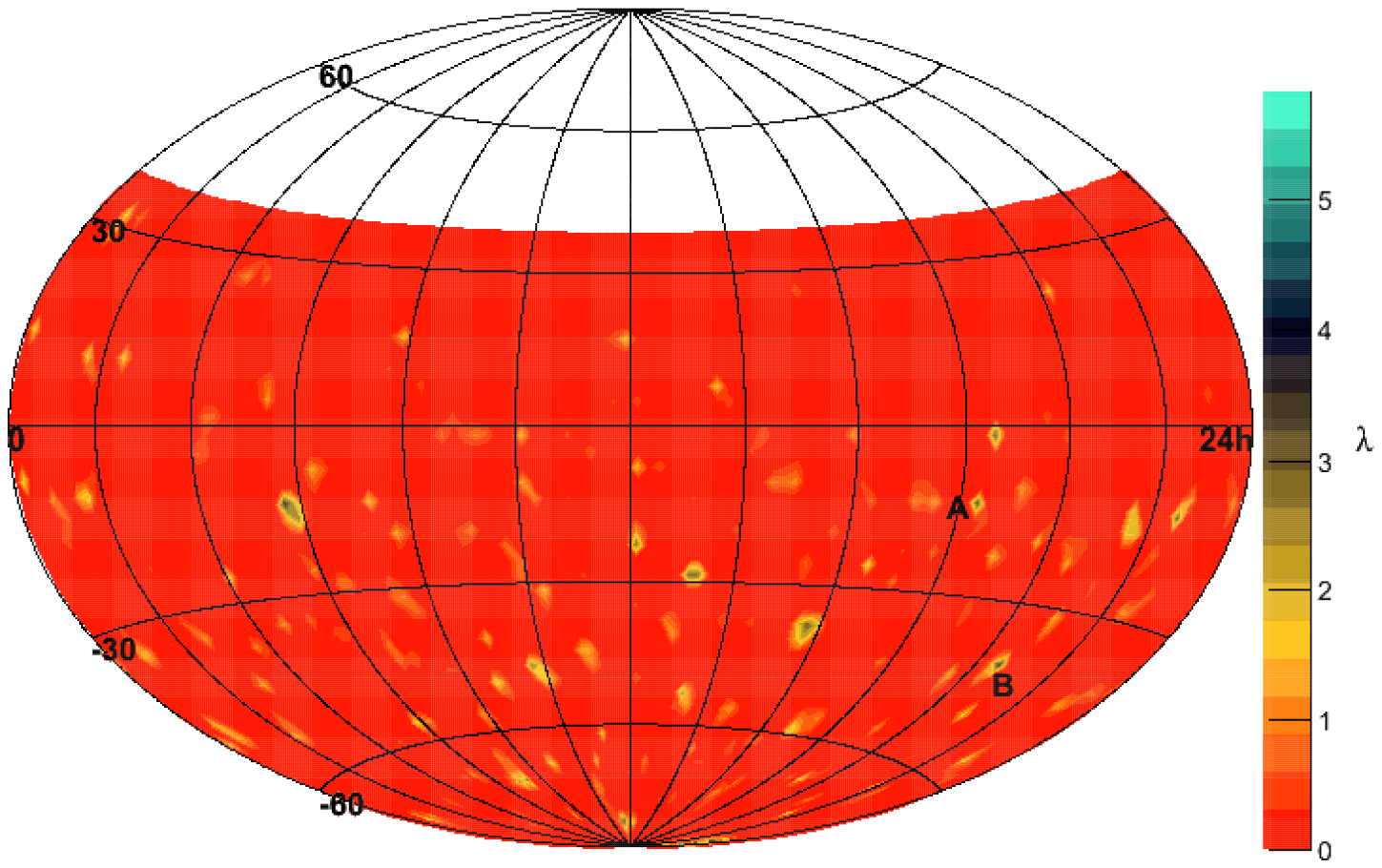} \\ }
\end{minipage}
\hfill%
\begin{minipage}[h]{0.53\linewidth}
\center{\includegraphics[width=1\linewidth]{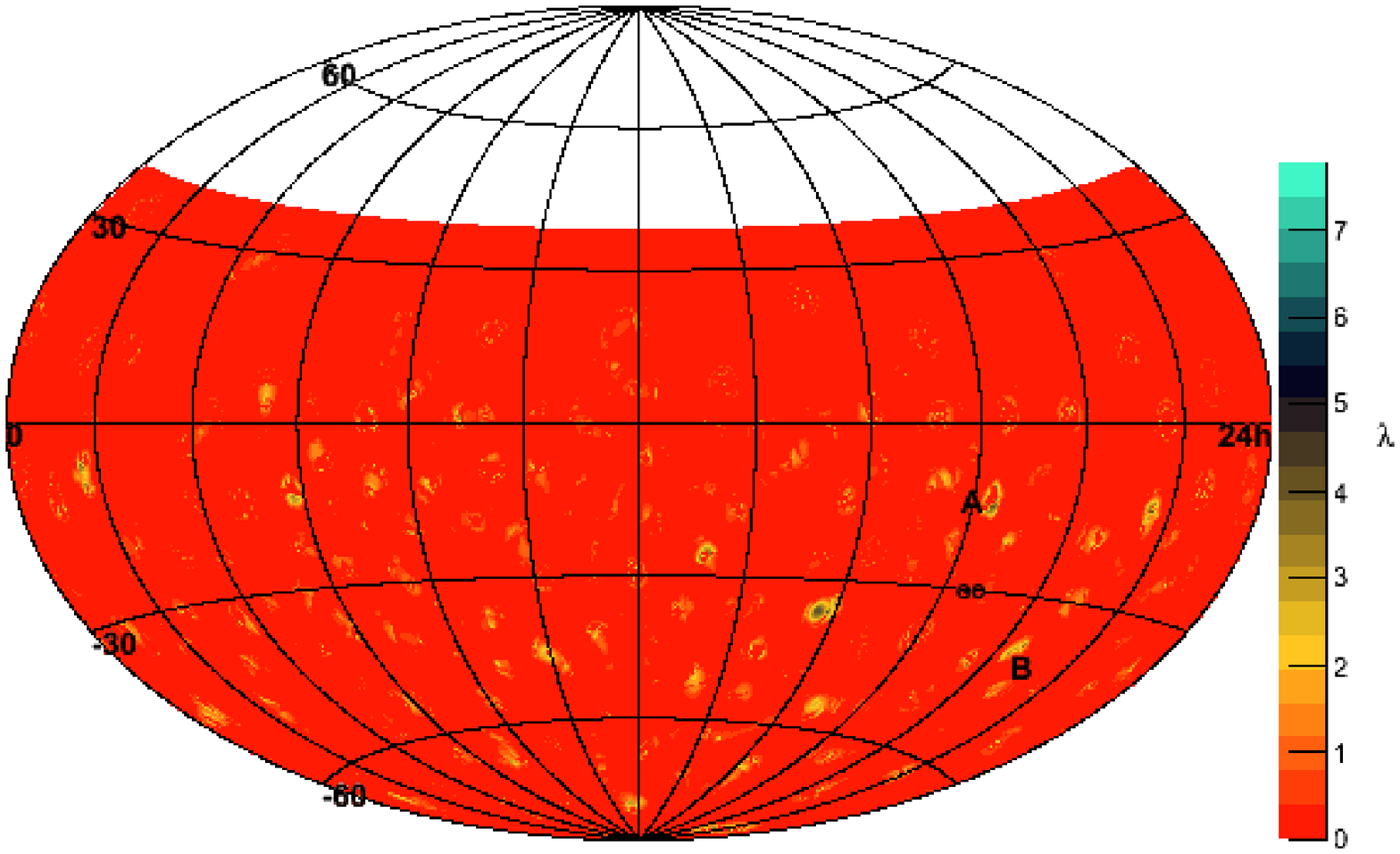} \\ }
\end{minipage}
\caption{\label{Qval396} Отношения функций правдоподобия $\lambda$ 
по данным НТ200 в небиновом методе сканирования неба в экваториальных координатах. 
Вероятные направления на скопление нейтринных событий: А -- $(18.3h,-13.25^{\circ})$
и В -- $(19.4h,-42.5^{\circ})$. Сканирование с шагом $3^{\circ}$ (слева) и $0.5^{\circ}$ (справа).}
\end{figure}

Результат сканирования направлений для НТ200 представлен на Рис.~\ref{Qval396} в экваториальных координатах  
с сеткой $ 3^\circ$ (слева) и с сеткой $ 0.5^\circ$ (справа) как распределения 
значений $\lambda$ с указанной шкалой цвета. 
\\

Направление, имеющее максимальное значение $\lambda$, соответствует наиболее
вероятному положению локального источника. Как видно из Рис.~\ref{Qval396} выделенными
оказываются три направления. При сканировании с шагом $3^\circ$ (Рис.~\ref{Qval396}, слева) 
точка B ($20.5h,-40.5^{\circ}$) с $\lambda= 3.5$ оказывается наболее вероятной, 
и этот вывод согласуется с результатами работы~\cite{NT200-AstroBelolap2013} 
бинового анализа по такому же набору нейтринных данных НТ200 (см. Рис.3 в ~\cite{NT200-AstroBelolap2013}). 
С шагом сканирования $ 0.5^\circ$ (Рис.~\ref{Qval396}, справа) направления локализованы точнее. 
Направлению В соответстуют координаты $(19.4h,-42.25^{\circ})$ и значение $\lambda= 5.3$.  Однако 
максимальное значение $\lambda=7.8$ относится к точке А с координатами $(18.3h,-13.25^{\circ})$, 
что совпадает с выводом нашего бинового сканирования (см.  Рис.~\ref{Z396}) и 
полученной наибольшей значимостью для направления А $(18.2h,-12^{\circ}$), 
как центра бина размером $4^\circ \times 4^\circ$. Исходя из предположения,
что количество обнаруженных событий имеют распределение Пуассона, рассчитывались значения
критериев согласия p-value. Для точки A величина p-value $= 0.21$, а для точки В величина p-value $= 0.34$,
то есть значимость для обоих точек меньше $1\sigma$. Для третьего выделенного направления с координатами  
$15.8h,-36.25^{\circ}$ значение $\lambda= 4.5$ и p-value $= 0.41$. Тем самым, так же как и в биновом анализе,
не обнаружено указаний на значимое скопление нейтринных событий на небесной сфере. Найденные
двумя методами вероятные направления одинаковые, но небиновый подход позволяет локализовать
их точнее.

\section{Верхний предел на поток нейтрино от предполагаемых кандидатов на нейтринные источники}\label{sec:levelUpL}
Аналогичный поиск небиновым методом был сделан для направлений предполагаемых 43-х источников, 
известных как гамма источники. Результат также не показал значимого превышения событий над фоном. 
Все выводы по этим источникам представлены в Таблице 1: координаты $\delta$-$\alpha$, 
число наблюдаемых и фоновых событий, доля временной видимости, значения $\lambda$ и p-value, 
верхние пределы на число сигнальных событий $s_{up}$ и поток мюонных нейтрино на $90\%$  доверительном уровне.

Верхний предел $s_{up}$ определялся методом Байеса как численное решение уравнения~(\ref{eq:CLup}) 
алгоритмом Брента в ROOTе. Кроме того, мы применили частотный подход нахождения верхних 
пределов, реализованный в моделях класса TRolke (ROOT), как с учетом систематических 
погрешностей, так и без них. Систематические ошибки
для НТ200 связаны с неопределенностью длины рассеяния света и чувствительностью 
оптических модулей. Как было показано ранее, например в работе~\cite{bBaiSyst}, для НТ200 
величина систематических погрешностей порядка $30\%$. 
 
Из полученных верхних пределов $s_{up}$ были сделаны оценки
верхних пределов на поток нейтрино в направлении на все исследуемые источники:
$$ \Phi_{\nu}^{up}=\frac{s^{up}}{\overline{S}_{eff} \cdot T},$$
где T -- живое время наблюдения с учетом $T_{vis}$ для каждого источника
и  $S_{eff}(E)$ -- эффективная площадь детектора, усредненная по нейтринному потоку
$\Phi_{\nu}(E)$ со спектром ускоренных нейтрино $E^{-2}$ от предполагаемого источника:
$$  \overline{S}_{eff}= \frac{\int S_{eff}(E)\cdot \Phi_{\nu}(E) dE}{\Phi_{\nu}(E)}.$$
В расчете предельных потоков для телескопа НТ200 использовалась эффективная площадь из работы~\cite{GRBBaikal}
и живое время наблюдения 1038 дней.
Для телескопа НТ1000 с пороговой энергией 1 ТэВ была сделана консервативная оценка чувствительности 
к потоку нейтрино от тех же источников 
за время наблюдения один год в предположении, что измерено нулевое число событий при угловом разрешении 
телескопа меньше градуса, а эффективная площадь соответствует $S_{eff}(E)$ из работы~\cite{GVD-NT1000} 
. 
\begin{figure}[h!]
\begin{minipage}[h]{0.53\linewidth}
\center{\includegraphics[width=1\linewidth]{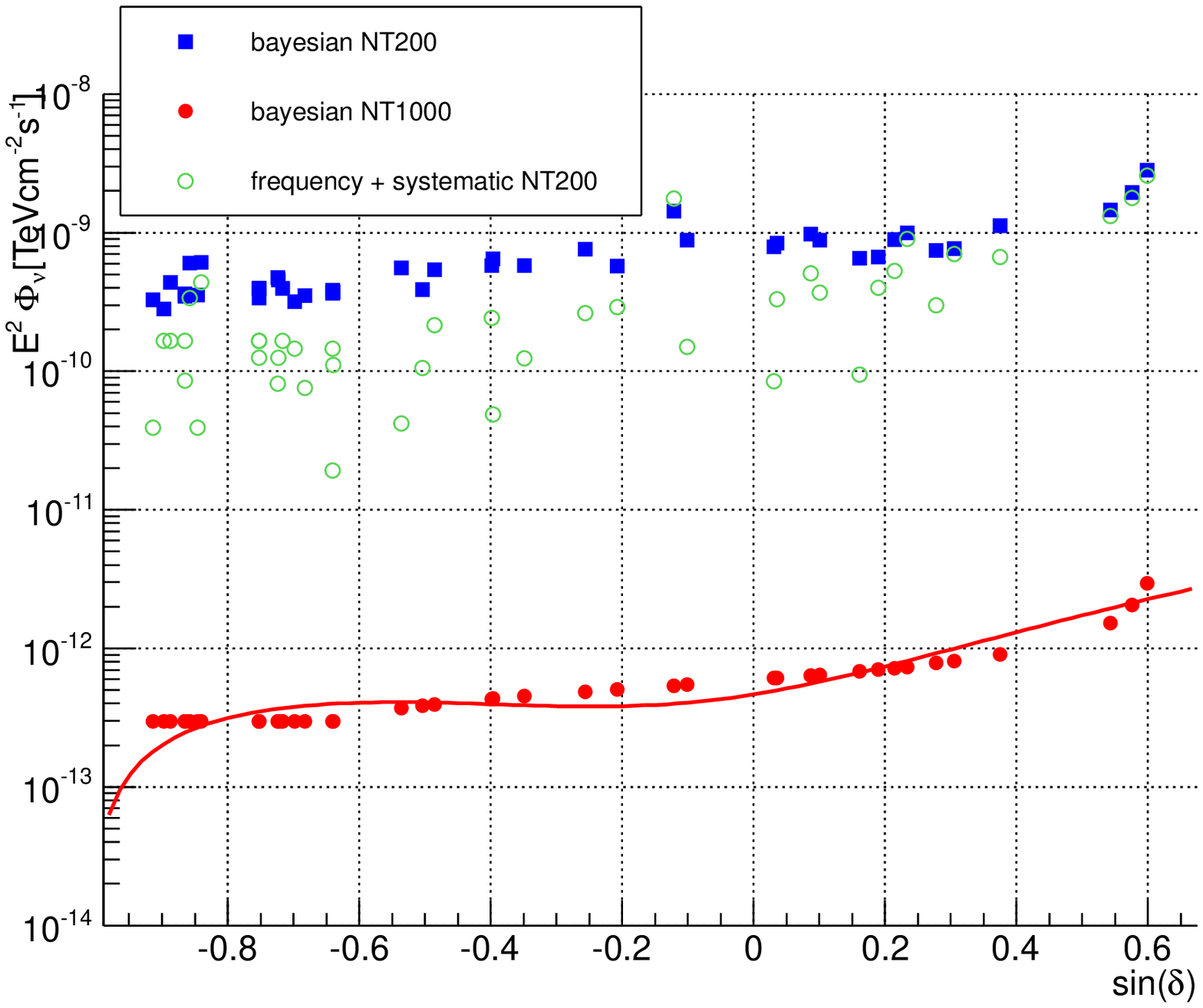} \\ }

\end{minipage}
\hfill%
\begin{minipage}[h]{0.53\linewidth}
\center{\includegraphics[width=1\linewidth]{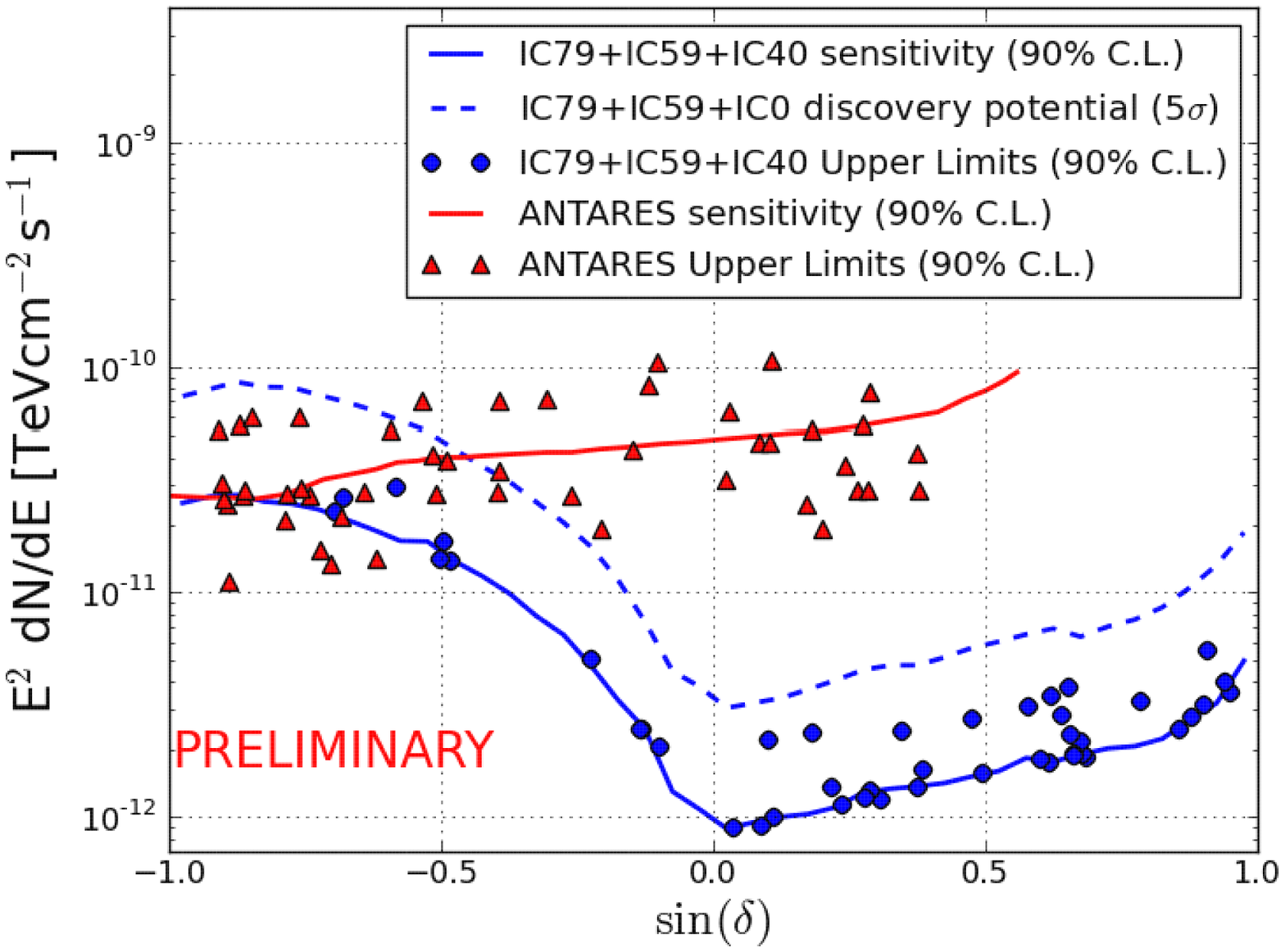} \\ }
\end{minipage}
\caption{\label{FlxLim} Слева: Верхние пределы на 90\% д.у. на поток
мюонных нейтрино от локальных астрофизических источников для телескопа
НТ200 и оценка чувствительности НТ1000 за год наблюдений в зависимости от
склонения (см. Таблицу 1). Справа: Верхние пределы на потоки мюонных
нейтрино для телескопов IceCube и ANTARES (см. ~\cite{IceCube-Lim2013}).}
\end{figure}\\



Полученные пределы на поток и чувствительность в зависимости от синуса склонения исследуемого источника 
показаны  на левом графике Рис.~\ref{FlxLim} для двух методов и с учетом систематики. Из сравнения всех полученных 
предельных значений по указанным обозначениям и фитирующим линиям следует, что наиболее консервативный результат 
дает метод Баейеса. Из сравнения с результатами других телескопов, ANTARES и IceCube,
показанных справа на Рис.~\ref{FlxLim} (график из работы~\cite{IceCube-Lim2013}) следует, что для НТ1000
чувствительность к локальным источникам за год наблюдений примерно на уровне предельных ограничений 
на поток телескопа IcеCube за время наблюдений в три раза большее.

\section{Заключение}\label{sec:levelSummary}
Исследовались статистические методы поиска скоплений событий или наиболее вероятных направлений 
источников нейтринно в Байкальском проекте глубоководных телескопов. Небиновым методом
локализация вероятных источников точнее. В результате сканирования небесной сферы и исследования 
направлений потенциальных источников 
указаний на значимое превышение измеренного числа событий над фоном не обнаружено. 
Для телескопа НТ200 за 1038 дней живого времени вычислены верхние ограничения на $90\%$ 
доверительном уровне на число сигнальных событий и на поток нейтрино от исследуемых 
астрофизических объектов. Для строящегося гигатонного телескопа НТ1000 получена оценка 
чувствительности к потоку нейтрино, как 
$$E_{\nu}^{-2}\cdot\Phi_{90\%} = (3 \div 30) \times 10^{-13}~\mbox{ТэВ}~\mbox{см}^{-2}~\mbox{c}^{-1},$$
из которой следует, что за год наблюдений уровень чувствительности НТ1000 к локальным источникам
сравним c предельными ограничениями на поток телескопа IcеCube за его примерно три года живого времени наблюдений. 
\begin{table}[H]
 \caption{Приложение 1. Таблица 1.}
 \tiny
 \begin{tabular}{|p{2cm}||p{0.6cm}||p{0.6cm}||p{0.5cm}||p{0.8cm}||p{0.5cm}||p{0.5cm}||p{0.6cm}||p{0.6cm}||p{2cm}||p{2.6cm}||p{2.05cm}|}

 \hline
 Source & $\alpha$ & $\delta $ & $\lambda$ & P-value & Sup& Nob & Nb & $T_{vis}$ &  $F200, TeV m^{-2}c^{-1}$ &$F200~+~sys, TeV m^{-2}c^{-1} $& $F1000, TeV m^{-2}c^{-1}$ \\
 \hline
$  3C~273$ & 187 & 2.05 & 0.122 & 0.97 & 7 & 3 & 3.32 & 0.486   & $8.37\times 10^{-10}$ & $3.3\times 10^{-10}$ & $6.09\times 10^{-13}$ \\
\hline
$ SS~433$ & 288 & 4.98 & 0 & 0.875 & 5 & 3 & 3.68 & 0.464   & $9.74\times 10^{-10}$ & $5.11\times 10^{-10}$ & $6.39\times 10^{-13}$ \\
\hline
$GRS~1915+105$ & 289 & 10.9 & 0 & 1 & 4 & 0 & 2.3 & 0.422 & $6.69\times 10^{-10}$ & $4.02\times 10^{-10}$ & $7.02\times 10^{-13}$ \\
\hline
$M87$ & 188 & 12.4 & 1.57 & 0.865 & 2 & 1 & 2.99 & 0.411 & $8.93\times 10^{-10}$ & $5.31\times 10^{-10}$ & $7.21\times 10^{-13}$ \\
\hline
$PKS~0528+134$ & 82.8 & 13.5 & 0 & 0.632 & 1 & 1 & 3.27 & 0.403 & $9.96\times 10^{-10}$ & $9.01\times 10^{-10}$ & $7.35\times 10^{-13}$ \\
\hline
$3C~454.3$ & 344 & 16.1 & 0 & 1 & 1 & 0 & 2.3 & 0.378 & $7.47\times 10^{-10}$ & $2.98\times 10^{-10}$ & $7.84\times 10^{-13}$ \\
\hline
$Geminga$ & 98.5 & 17.8 & 0 & 1 & 0 & 0 & 2.3 & 0.367 & $7.7\times 10^{-10}$ & $7\times 10^{-10}$ & $8.08\times 10^{-13}$ \\
\hline
$Crab~Nebula$ & 83.7 & 22 & 0.557 & 0.865 & 2 & 1 & 2.99 & 0.328 & $1.12e-09$ & $6.67\times 10^{-10}$ & $9.04\times 10^{-13}$ \\
\hline
$GRO~J0422+32$ & 65.4 & 32.9 & 0 & 1 & 0 & 0 & 2.3 & 0.194 & $1.45e-09$ & $1.32e-09$ & $1.52\times 10^{-12}$ \\
\hline
$Cyg~X-1$ & 300 & 35.2 & 0 & 1 & 0 & 0 & 2.3 & 0.144 & $1.95e-09$ & $1.78e-09$ & $2.05\times 10^{-12}$ \\
\hline
$MGRO~J2019+37$ & 305 & 36.8 & 0 & 1 & 0 & 0 & 2.3 & 0.1 & $2.82e-09$ & $2.57e-09$ & $2.96\times 10^{-12}$ \\
\hline
$SGR~1900+14$ & 287 & 9.3 & 0 & 1 & 5 & 0 & 2.3 & 0.433 & $6.51\times 10^{-10}$ & $9.45\times 10^{-11}$ & $6.84\times 10^{-13}$ \\
\hline
$SGR~0526-66$ & 81.5 & -66 & 0 & 1 & 12 & 2 & 2.67 & 1 & $3.28\times 10^{-10}$ & $3.9\times 10^{-11}$ & $2.96\times 10^{-13}$ \\
\hline
$1E~1048.1-5937$ & 162 & -59.9 & 0 & 0.999 & 11 & 3 & 2.96 & 1 & $3.62\times 10^{-10}$ & $1.66\times 10^{-10}$ & $2.96\times 10^{-13}$ \\
\hline
$SGR~1806-20$ & 272 & -20.4 & 0 & 0.994 & 9 & 3 & 3.1 & 0.656 & $5.8\times 10^{-10}$ & $1.24\times 10^{-10}$ & $4.52\times 10^{-13}$ \\
\hline
$Vela~X$ & 128 & -45.8 & 0.0126 & 0.995 & 11 & 4 & 3.24 & 1 & $3.97\times 10^{-10}$ & $1.66\times 10^{-10}$ & $2.96\times 10^{-13}$ \\
\hline
$G~343.1-2.3$ & 257 & -44.3 & 0 & 1 & 15 & 2 & 2.6 & 1 & $3.19\times 10^{-10}$ & $1.45\times 10^{-10}$ & $2.96\times 10^{-13}$ \\
\hline
$MSH~15-52$ & 228 & -59.1 & 0 & 0.857 & 11 & 8 & 4.9 & 1 & $6.01\times 10^{-10}$ & $3.37\times 10^{-10}$ & $2.96\times 10^{-13}$ \\
\hline
$Vela~Jr$ & 133 & -46.3 & 0.701 & 0.971 & 10 & 5 & 3.71 & 1 & $4.55\times 10^{-10}$ & $1.25\times 10^{-10}$ & $2.96\times 10^{-13}$ \\
\hline
$GRO~J1655-40$ & 254 & -39.8 & 0 & 1 & 16 & 5 & 3.14 & 1 & $3.85\times 10^{-10}$ & $1.93\times 10^{-11}$ & $2.96\times 10^{-13}$ \\
\hline
$HESS~J1023-575$ & 156 & -57.8 & 0 & 0.999 & 12 & 3 & 2.9 & 1 & $3.56\times 10^{-10}$ & $3.9\times 10^{-11}$ & $2.96\times 10^{-13}$ \\
\hline
$GX~339$ & 256 & -48.8 & 0 & 0.995 & 11 & 4 & 3.24 & 1 & $3.97\times 10^{-10}$ & $1.66\times 10^{-10}$ & $2.96\times 10^{-13}$ \\
\hline
$RX~J1713.7-3946$ & 258 & -39.8 & 0 & 1 & 14 & 4 & 3.03 & 1 & $3.71\times 10^{-10}$ & $1.11\times 10^{-10}$ & $2.96\times 10^{-13}$ \\
\hline
$HESS~J1837-069$ & 279 & -6.95 & 0 & 0.185 & 3 & 5 & 6.44 & 0.55 & $1.43e-09$ & $1.76e-09$ & $5.39\times 10^{-13}$ \\
\hline
$1ES~0346-121$ & 57.4 & -12 & 0 & 0.982 & 4 & 1 & 2.74 & 0.586 & $5.73\times 10^{-10}$ & $2.9\times 10^{-10}$ & $5.05\times 10^{-13}$ \\
\hline
$3C~279$ & 194 & -5.79 & 0 & 0.945 & 9 & 5 & 3.88 & 0.542 & $8.79\times 10^{-10}$ & $1.5\times 10^{-10}$ & $5.47\times 10^{-13}$ \\
\hline
$Cir~X-1$ & 230 & -57.2 & 0 & 0.793 & 9 & 7 & 4.96 & 1 & $6.08\times 10^{-10}$ & $4.36\times 10^{-10}$ & $2.96\times 10^{-13}$ \\
\hline
$PKS~2005-489$ & 302 & -48.8 & 0.234 & 1 & 10 & 2 & 2.75 & 1 & $3.37\times 10^{-10}$ & $1.25\times 10^{-10}$ & $2.96\times 10^{-13}$ \\
\hline
$GC$ & 266 & -29 & 0 & 0.97 & 7 & 3 & 3.32 & 0.75 & $5.42\times 10^{-10}$ & $2.14\times 10^{-10}$ & $3.95\times 10^{-13}$ \\
\hline
$LS~5039$ & 277 & -14.8 & 0 & 0.918 & 7 & 4 & 3.8 & 0.611 & $7.62\times 10^{-10}$ & $2.63\times 10^{-10}$ & $4.85\times 10^{-13}$ \\
\hline
$RX~J085.0-4622$ & 133 & -46.4 & 0.607 & 0.945 & 9 & 5 & 3.88 & 1 & $4.76\times 10^{-10}$ & $8.1\times 10^{-11}$ & $2.96\times 10^{-13}$ \\
\hline
$PKS~0548-322$ & 87.7 & -32.4 & 0 & 0.958 & 8 & 4 & 3.61 & 0.797 & $5.55\times 10^{-10}$ & $4.18\times 10^{-11}$ & $3.72\times 10^{-13}$ \\
\hline
$PSR~B1259-63$ & 196 & -63.8 & 0 & 1 & 11 & 0 & 2.3 & 1 & $2.82\times 10^{-10}$ & $1.66\times 10^{-10}$ & $2.96\times 10^{-13}$ \\
\hline
$PKS~2155-304$ & 330 & -30.2 & 0 & 1 & 18 & 1 & 2.42 & 0.764 & $3.89\times 10^{-10}$ & $1.05\times 10^{-10}$ & $3.88\times 10^{-13}$ \\
\hline
$HESS~J0632+057$ & 98.2 & 5.81 & 0.7 & 0.908 & 4 & 2 & 3.29 & 0.458 & $8.8\times 10^{-10}$ & $3.7\times 10^{-10}$ & $6.46\times 10^{-13}$ \\
\hline
$RCW~86$ & 221 & -62.5 & 0.982 & 0.985 & 11 & 5 & 3.57 & 1 & $4.38\times 10^{-10}$ & $1.66\times 10^{-10}$ & $2.96\times 10^{-13}$ \\
\hline
$RGB~J0152+017$ & 28.2 & 1.79 & 1.9 & 0.96 & 5 & 2 & 3.13 & 0.486 & $7.88\times 10^{-10}$ & $8.43\times 10^{-11}$ & $6.09\times 10^{-13}$ \\
\hline
$Cen~A$ & 201 & -43 & 0 & 1 & 13 & 3 & 2.85 & 1 & $3.5\times 10^{-10}$ & $7.6\times 10^{-11}$ & $2.96\times 10^{-13}$ \\
\hline
$ESO~139-G12$ & 264 & -59.9 & 0 & 0.95 & 3 & 1 & 2.84 & 1 & $3.48\times 10^{-10}$ & $8.52\times 10^{-11}$ & $2.96\times 10^{-13}$ \\
\hline
$W28$ & 270 & -23.3 & 0 & 0.958 & 8 & 4 & 3.61 & 0.683 & $6.48\times 10^{-10}$ & $4.88\times 10^{-11}$ & $4.34\times 10^{-13}$ \\
\hline
$1ES~1101-232$ & 166 & -23.5 & 0 & 0.995 & 11 & 4 & 3.24 & 0.686 & $5.79\times 10^{-10}$ & $2.41\times 10^{-10}$ & $4.32\times 10^{-13}$ \\
\hline
\end{tabular}
\end{table}

\end{document}